# Construction of a Digital Hadron Calorimeter

José Repond[1]

For the DHCAL Collaboration

1 – Argonne National Laboratory, 9700 S. Cass Avenue, Argonne, IL, U.S.A.

The DHCAL collaboration is assembling a large scale prototype Digital Hadron Calorimeter (DHCAL). The calorimeter utilizes Resistive Plate Chambers (RPCs) as active medium and features of the order of 400,000 1×1 cm$^2$ pads with binary (or digital) electronic readout. The purpose of the prototype is to provide detailed measurements of hadronic showers and to prove the concept of a DHCAL with RPCs as active elements.

## 1   Particle Flow Algorithms and Digital Hadron Calorimeter

To fully exploit the physics potential of a future lepton collider, operating at center-of-mass energies of 0.5 TeV and above, requires an unprecedented jet energy resolution, of the order of a factor of two better than previously attained. The preferred approach to attain this performance is through the application of Particle Flow Algorithms (PFAs) [1], which utilize the tracker to measure the momenta of charged particle, the electromagnetic calorimeter to measure the energy of photons and the combined electromagnetic and hadronic calorimeters to measure the energy of neutral hadrons, i.e. the neutrons and $K_L^0$'s. The energy of jets is then reconstructed combining these measurements into one. The major challenge of PFAs is the association of energy deposits in the calorimeter to the charged or neutral particles originating from the interaction point. This challenge is met with the development of highly segmented or so-called imaging calorimeters [2].

In this context we present the status of the construction of a prototype Digital Hadron Calorimeter (DHCAL) utilizing Resistive Plate Chambers (RPCs) as active medium. The chambers are read out with 1×1 cm$^2$ pads and a binary (or digital) electronic readout system [3]. The latter is based on the DCAL chip [4], which was developed by Argonne and Fermilab, and which is optimized for the readout of large number of channels.

The project is being carried out by the DHCAL collaboration [5], including 36 people from Argonne National Laboratory, Boston University, Fermilab, Iowa University and the University of Texas at Arlington. The testing of the prototype calorimeter will be part of the scientific program of the international CALICE collaboration [6], which develops finely segmented calorimetry for future lepton colliders.

## 2   The 1 m$^3$ Prototype Calorimeter

The prototype DHCAL will contain 38 layers, each with an area of 96×96 cm$^2$. The layers will be inserted into the existing absorber structure of the Scintillator-Steel prototype hadron calorimeter [7] of the CALICE collaboration. The absorber plates measure 100×100×1.6 cm$^2$ each and correspond in thickness to about 0.9 radiation lengths $X_0$ or to 0.1 interaction lengths $\lambda_I$. Each detector layer will contain three RPCs with glass as resistive plates. The thicknesses of the glass are 1.15 mm and 0.85 mm for the cathode and anode, respectively. The use of thinner glass on the anode side is to reduce the pad multiplicity [8]. The gas gap between the plates is maintained with fishing lines with a diameter of 1.15 mm.

Each chamber is read out by two readout boards, each covering an area of 48 x 32 cm$^2$ and each containing a pad board and a front-end board with the DCAL ASICs. The connection between the two boards is provided by drops of conductive glue.

In the following we briefly review the status of the various tasks of the construction and assembly effort.

## 3  RPC Construction

The RPC construction involves the following steps:

- Spraying a resistive coat onto one side of the glass plates
- Cutting the frame pieces using a specially designed fixture
- Gluing the frame pieces into a frame using a special gluing fixture
- Gluing the thick glass plate onto the frame
- Stringing the fishing lines
- Gluing the thin glass plate onto the other side of the frame
- Taping sheets of Mylar to insolate and protect the resistive coats
- Installing of the high voltage cable and connector.

The spraying of the glass plates proved to be a challenging task. To obtain a reasonably low pad multiplicity (of the order of 1.4 for 90% detection efficiency), the surface resistivity of the resistive paint of the anode plate is required to be in the range of 1 to 5 MΩ/□. In comparison, the values for the cathode plates are not as critical. Figure 1 shows the surface resistivity for the plates sprayed so far. The error bars indicate the root-mean-square of the values across the surface of the plates. After plate number 130 a large improvement was achieved by using a commercial spraying gun rather than an artist airbrush. For the later plates, the root-mean-square of the values across the surface is approximately 25% of the average value of a given plate. The current yield is close to 100%.

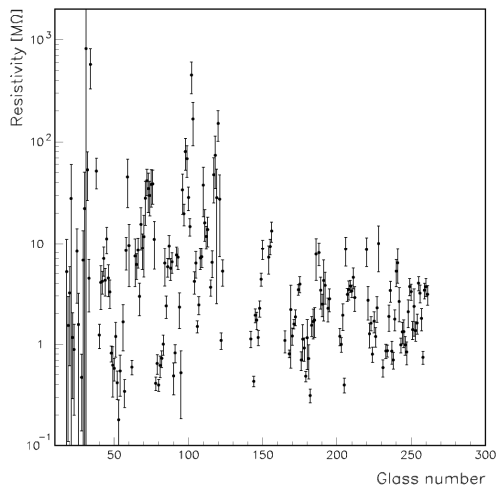

Figure 1: Surface resistivity for the glass plates sprayed to date. The error bars indicate the root-mean-square of the values over the surface of the plates.

The assembly of the chambers turned out to be challenging as well. Particular care is devoted to maintaining a constant thickness of the chambers, thus minimizing variations in response over the area of the chamber. Measurements of the thickness around the perimeter show a root-mean-square of approximately 2%, dominated by the corners which in general are somewhat thicker. As the glass bends easily, the thickness of the chambers away from the edges is difficult to measure. However, it is assumed that here the fishing lines provide a uniform gap. Figure 2 shows a stack of completed chambers (not all of them equipped with the high voltage connector).

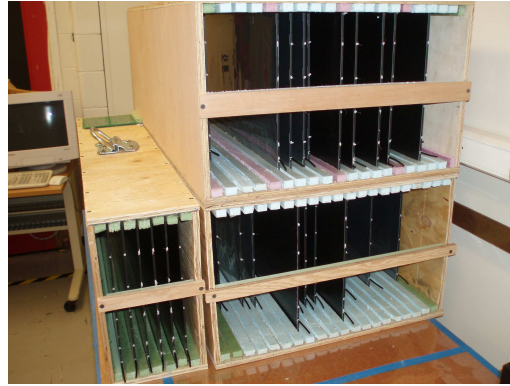

Figure 2: Photograph of completed RPCs (some with the high voltage connector not yet installed).

Of the 114 chambers needed to fully instrument the prototype calorimeter, about 80 have been built to date. Cosmic ray testing of the completed chambers has begun and shows the expected performance.

## 4  Cassette Assembly

Three completed RPCs are inserted into a cassette structure consisting of a 2 mm stainless steel back plane and a 2 mm Copper top plate. The purpose of the cassettes is to protect the chambers, cool the front-end electronics and to slightly press the readout board onto the anode glass plate, to eliminate possible air gaps between the two. Figure 3 shows a photograph of a completed cassette, where part of the electronics boards are seen to protrude beyond the area of the copper sheets. .

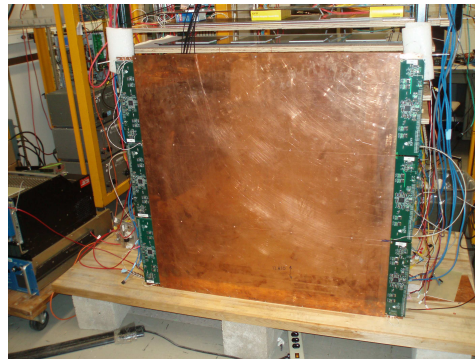

Figure 3: Assembled detector cassette.

Having subdivided the plane into three chambers provides the possibility to insert a string between the chambers to exert a small compression of the plane. Tests with various nylon strings were successful and provided the necessary force.

A small copper tube on top of the cassette and connected to a water chiller will provide the necessary cooling of the front-end electronics. Each plane is expected to generate approximately 100 Watts.

## 5    Peripherals

The RPCs are run with a mixture of three gases: Freon R134a (94.5%), Isobutan (5.0%) and Sulfur-hexafluoride (0.5%). A mixing station with electronic controls of the mixing fractions has been designed, built and tested. A distribution rack will fan the mixed gas into 25 lines (with individual flow regulators and bubblers). Each line will feed six chambers or two layers. The distribution rack has been assembled and tested.

The front-end electronics operate on +5V. The power will be supplied by seven Wiener power supplies (in hand) and will be distributed to the individual boards through custom distribution boxes, see Figure 4. Each such box will supply power to the boards of eight detector layers.

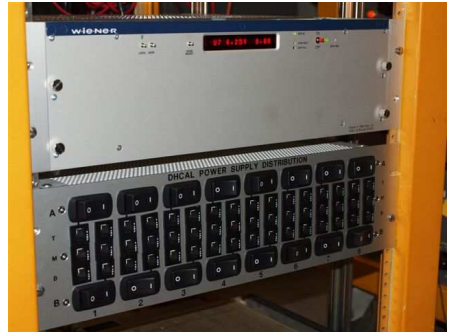

Figure 4: Photograph of the Wiener low voltage power supply and the distribution box.

Our RPCs typically run with a High Voltage (HV) around 6.2 kV. The power will be supplied by LeCroy 4032 units. A computer program to control the voltage setting and monitor the values has been developed.

## 6    Readout System Overview

The electronic readout system [3] was developed around the so-called DCAL chip [4]. Each DCAL chip connects to 64 channels or pads and applies a single threshold (corresponding to a binary or digital readout). The chip provides a timestamp (with a time resolution of 100 ns) and the hit pattern of the 64 channels. The chip operates either in high gain (for application with GEMs) or in low gain mode (optimized for RPCs) and in triggered (for internal charge injection, cosmic rays and test beam) or trigger-less mode (for noise measurements). The chips are mounted on front-end boards, where six such boards complete the readout of a detector layer, see Figure 5.

Figure 5: Photograph of a detector layer with six front-end boards.

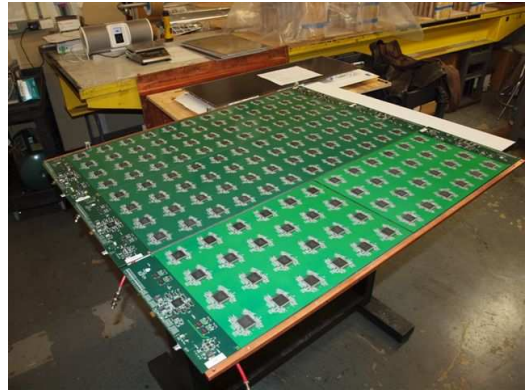

The pad-boards are separate entities from the front-end board. The connections between the two are provided by drops of conductive glue. The signals from the 24 DCAL chips on a given front-end board are routed to the edge of the board where a Field-Programmable Gate-Array (FPGA) receives the data, applies zero-suppression and routes the information to the next level, the Data Collector (DCOL) modules. The DCOLs are VME modules and process data from up to twelve front-end boards each. A Timing and Trigger Module (TTM) coordinates the timings and distributes the clock signals, resets, slow controls and triggers via the DCOLs to the front-end.

## 7   Construction of the Readout system

The DCAL chips were produced in an engineering run, yielding 11 wafers containing about 10,300 chips. A small number of chips were tested extensively on the bench at Argonne. As an example, Figure 6 shows the S-curves for the 64 channels of a chip. As the threshold is lowered the chip goes from recording zero hits to recording all hits corresponding to the 100 test charges (internally) injected into the front-end. Note the excellent uniformity from channel to channel. Overall not a single design fault of the DCAL was discovered in these detailed bench tests. Robotic testing at Fermilab identified 8644 good parts (5472 chips are needed to fully equip the prototype calorimeter).

Figure 6: S-Curve for the 64 channels of a DCAL chip. For each threshold setting 100 test charges are injected into the front-end.

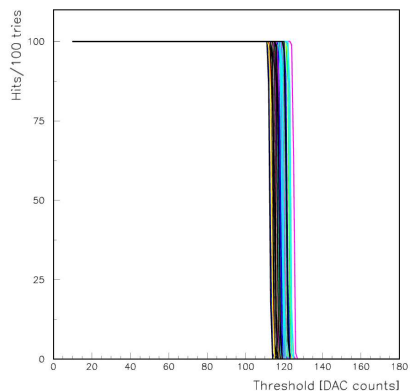

The design of the front-end board underwent several iterations. Particular care was devoted to minimize the cross-talk from the digital lines onto the analog inputs. Measurements of the electronic noise floor (with the detector high voltage turned off) show values of less than 5 fC in high gain mode. In addition, the boards underwent extensive torture tests and appear to perform reliably and error free. The boards are being fabricated and production assembly is about to start.

A gluing fixture to dispense the conductive glue dots onto the back-side of the pad board has been built. Figure 7 shows a photograph taken while the last dots are being applied to a pad board. To date the fixture glued nine boards, all successfully. The dispensing of the glue dots requires 55 minutes per board.

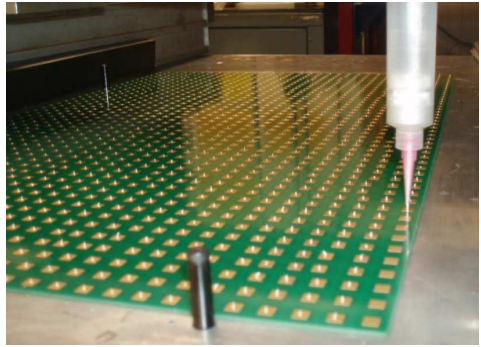

Figure 7: Photograph of a pad-board, showing the gluing fixture in the process of applying the last dots of conductive glue.

The DCOLs have been produced and their testing is close to complete. The TTMs required to be redesigned in order to implement a master-slave feature, which enables the use of multiple VME crates. The re-designed TTMs are being assembled.

## 8 Software

For compatibility reasons the data acquisition software is based on the CALICE online software. It has been written, debugged and been used extensively for several years.

A new offline event builder is being developed. This is a somewhat challenging task as the data from different DCAL chips is a priori not ordered in time. The event builder also must be able to handle events or records with errors, such as caused by for instance dropping a bit.

The setup has been simulated by a GEANT4 based simulation together with a standalone program emulating the response of RPCs [9]. The predicted resolution for pions with energies up to 28 GeV is approximately $58\%/\sqrt{E}$ with a negligible constant term [10,11].

The CALICE event display program has been adapted to the geometry of the prototype calorimeter. Figure 8 shows an event containing a simulated 60 GeV pion shower.

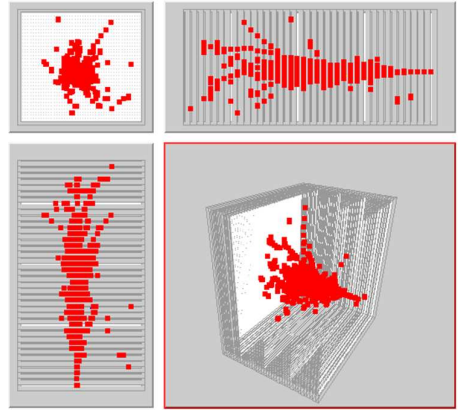

Figure 8: Event display of a simulated 60 GeV pion shower in the prototype calorimeter.

## 9   Time Schedule and Plans

In conclusion Table 1 summarizes the time schedule for the remaining construction jobs and the subsequent testing in the Fermilab test beam [12].

| Task | Dates | Comments |
| --- | --- | --- |
| Construction | Complete by June 30th | Should not slip much more… |
| Cosmic ray testing of cubic meter | April through August | |
| Installation into Mtest | September | |
| 1st data taking period | October | DHCAL standalone (with TCMT) |
| 2nd data taking period | December | Combined with ECAL |
| 3rd data taking period | Early in 2011 | DHCAL standalone or combined |
| Disassembly and shipping of the stage | March 2011 | (Maybe not so) hard deadline |

Table 1: Time schedule for the completion of the construction and the testing in the Fermilab testbeam.

## 10   Acknowledgements

The author would like to thank the organizers for an impeccably organized workshop and for the opportunity to present the work of the DHCAL collaboration to the International Linear Collider community.